\documentclass[journal]{rmaa}

\usepackage{epstopdf}
\usepackage{textcomp}
\usepackage{multirow} 

\title{New Radio Continuum Observations of the Compact Source Projected Inside NGC~6334A}

 \author{
Luis F. Rodr\'\i guez\altaffilmark{1,2},
Josep M. Masqu\'e\altaffilmark{1},
Sergio A. Dzib\altaffilmark{1},
Laurent Loinard\altaffilmark{1},
and
Stanley E. Kurtz\altaffilmark{1}}
  \altaffiltext{1}{Centro de Radioastronom\'{\i}a y Astrof\'{\i}sica, Universidad
Nacional Aut\'onoma de M\'exico, Campus Morelia} 

\altaffiltext{2}{Astronomy Department, Faculty of Science, King Abdulaziz University, P.O.
Box 80203, Jeddah 21589, Saudi Arabia}

\fulladdresses{
\item Sergio A. Dzib, Stanley E. Kurtz,
Laurent Loinard, Josep M. Masqu\'e, Luis F. Rodr\'\i guez: Centro de Radiostronom\'{\i}a
y Astrof\'\i sica, 
UNAM, A. P. 3-72, 
(Xangari), 58089 Morelia, Michoac\'an, M\'exico
(s.dzib, s.kurtz, l.loinard, j.masque, l.rodriguez@crya.unam.mx).
}

\shortauthor{Rodr\'\i guez et al.}
\shorttitle{Radio Continuum Emission from FS CMa Stars}

\SetVolume{50} \SetFirstPage{001} \SetYear{2013}

\resumen{
Se sabe que unas pocas regiones HII muestran una fuente compacta de radio cerca de su
centro. La naturaleza de estas fuentes compactas no est\'a bien
establecida. Presentamos el an\'alisis de datos tanto nuevos como de archivo del
Very Large Array de la fuente compacta proyectada cerca del centro de
la regi\'on HII NGC~6334A, parte del complejo de NGC~6334. Mostramos que
la fuente es variable en el tiempo en escalas de a\~nos y para una \'epoca determinamos
un espectro no-t\'ermico, sugiriendo emisi\'on sincrotr\'onica.
Proponemos que esta fuente es la regi\'on de interacci\'on de los vientos de
un sistema binario masivo que podr\'\i a ser la fuente ionizadora de NGC~6334A.
}
 
\abstract{
A handful of HII regions are known to exhibit a compact radio source near their centers.
The nature of these compact radio sources is not well
established. We present the analysis of new as well as archival Very Large
Array observations of the compact source projected near the center
of the NGC~6334A HII region, part of the NGC~6334 complex.
We show that the compact source is time variable on a scale of years and
determine for one epoch a non-thermal spectrum, suggestive of
synchrotron emission. We propose that this source could be the wind interaction region of
a massive binary system that could be the ionizing source of NGC~6334A.
}
      
\keywords{ISM: HII regions --- ISM: individual (NGC 6334A) --- RADIO CONTINUUM: STARS}

\begin{document}

\maketitle

\section{Introduction}

It is well known that the centimeter continuum radiation from classic HII regions is dominated by 
strong free-free emission 
from the extended ionized gas present there. However, when observed with the high angular
resolution provided by an interferometer, the extended emission is filtered out and one starts
to detect compact sub-arcsecond sources of various natures. These sources include
thermal (i.e. free-free) emitters such as 
hypercompact (HC) HII regions, externally ionized globules, proplyds,
thermal jets, and ionized stellar winds. On the other hand, non thermal emitters include
young low-mass stars that can have strong magnetospheric activity and emit detectable
gyrosynchrotron radiation
and massive binary stars that can produce synchrotron radiation in the 
region where their winds collide (see Rodr\'\i guez et al. 2012a for a
more detailed discussion on these different types of radio sources).

Evidence for what could be a new type of compact radio source has recently become available.
In the case of the well studied ultracompact HII region W3(OH),
Dzib et al. (2013a) have reported that a compact radio source at its center is time variable
and has a positive spectral index and brightness temperature suggestive of
partially optically-thick free-free radiation. This source is associated with
the exciting source of the HII region and Dzib et al. (2013a)
tentatively propose that the
radio emission could arise in a static ionized atmosphere around
a fossil photoevaporated disk. This radio source was originally detected 
by Kawamura \& Masson (1998) but no further research has been published.

A similar compact radio source was reported by Carral et al. (2002), in this case at the center of
the compact HII region NGC~6334A.
This region, first reported by Rodr\'\i guez, Cant\'o \& Moran (1982), is found in the
massive star-forming complex NGC 6334, at 1.7 kpc of distance. The
location of the compact radio source, centered in the shell-like morphology
displayed by NGC 6334A (see Figure 4 of Carral et al. 2002), suggests an association with the star exciting the H II
region. To obtain additional information on the nature 
of these continuum radio sources at the center of compact
and ultracompact HII regions, we have analyzed new as well as
unpublished high angular resolution archive data from the Very
Large Array (VLA)
of the NRAO\footnote{The National Radio
Astronomy Observatory is operated by Associated Universities
Inc. under cooperative agreement with the National Science Foundation.}
taken toward NGC~6334A.

\begin{table*}[htbp]
\footnotesize
  \setlength{\tabnotewidth}{1.8\columnwidth} 
    \tablecols{8} 
\small
  \caption{Very Large Array Observations}
    \begin{center}
	\begin{tabular}{lccccccc}\hline\hline
	 &  Frequency & Phase  & Bootstrapped & & Synthesized & Flux Compact &  \\ 
	Epoch &  (GHz) & Calibrator  & Flux (Jy) & Conf. & Beam$^{b}$ & Source (mJy) & Project \\ 
	\hline
	1997 Feb 02 & 8.46  &  1733$-$130 & 9.18$\pm$0.01  & A & $0\rlap.{''}67\times0\rlap.{''}53; +29^\circ$
	& 7.0$\pm$0.8 &  AT202 \\
	2002 May 01 & 8.46  &  1744$-$312 & 0.615$\pm$0.001 & A & $0\rlap.{''}53\times0\rlap.{''}20; -12^\circ$
	& 3.3$\pm$0.5 & AR474 \\
	2002 May 01 & 14.9  &  1744$-$312 & 0.860$\pm$0.009 & A & $0\rlap.{''}30\times0\rlap.{''}12; +1^\circ$
        & 1.7$\pm$0.4 & AR474 \\
	2002 May 01 & 43.3  &  1744$-$312 & 1.00$\pm$0.01 & A & $0\rlap.{''}15\times0\rlap.{''}05; -7^\circ$
	& $\leq$1.0$^a$ & AR474 \\
	2006 Aug 18 & 8.46  &  1744$-$312 & 0.587$\pm$0.005 & B & $1\rlap.{''}94\times0\rlap.{''}44; +3^\circ$
	& $\leq$2.9$^a$ &  S7810 \\
	\hline\hline
\tabnotetext{a}{Three-sigma upper limit.}	
\tabnotetext{b}{Half power full width dimensions in arc seconds and position of major axis in
degrees.}
  \label{tab:1}
\end{tabular}
\end{center}
\end{table*}

\begin{figure}
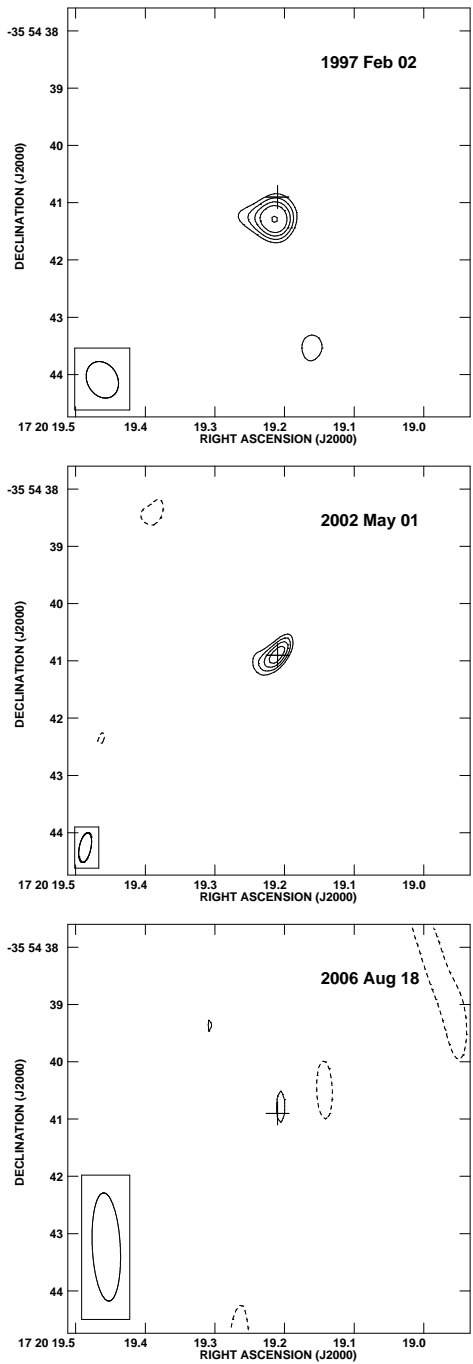

\centering
\includegraphics[scale=0.35, angle=0]{N6334XX97.PS}
\includegraphics[scale=0.35, angle=0]{N6334XX02.PS}
\includegraphics[scale=0.35, angle=0]{N6334XX06.PS}

 \caption{VLA contour images of the 8.46 GHz continuum emission toward
the compact source at the center of NGC~6334A for three different epochs. 
Contours are -4, -3, 3, 4, 5, 6, and 8
times the rms noise of each image (840, 450 and 960 $\mu$Jy beam$^{-1}$ for the 1997,
2002, and 2006 images, respectively). The cross marks the position of the compact source
derived from the 8.46 GHz image of 2002, $\alpha(2000) = 17^h~20^m~19\rlap.{''}21;~
\delta(2000) = -35^\circ~54'~40\rlap.{''}9$.
}
  \label{fig1}
\end{figure}

\begin{figure}
\centering
\includegraphics[scale=0.4, angle=0]{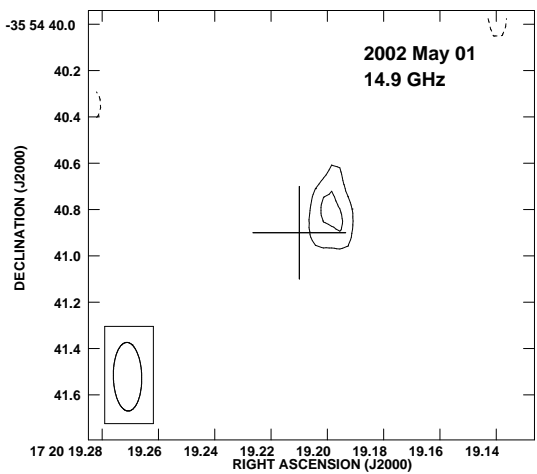}
 \caption{VLA contour image of the 14.9 GHz continuum emission toward
 the compact source in NGC~6334A.
 Contours are -3, 3, and 4
 times 400 $\mu$Jy beam$^{-1}$, the rms noise of the image.
 The cross marks the position of the compact source
 derived from the 8.46 GHz image of the same epoch.
 }
   \label{fig2}
 \end{figure}

\section{Data Reduction}

The search for these compact radio sources embedded in larger structures can only
be performed using observations of the highest angular resolution possible.
This seriously limits the observations that can be used.
In Table 1 we list the data used in this study. The 1997 observations are those
already reported by Carral et al. (2002). The 2002 observations were taken by us as a follow up
study to the Carral et al. results. Finally, the 2006 data are available in the VLA archives.
The data were edited and calibrated using the software package Astronomical Image
Processing System (AIPS) of NRAO. The 8.46 GHz observations were
self-calibrated in phase and amplitude. The images were made using only
visibilities with a baseline larger than 75 k$\lambda$, suppressing
structures larger than $\sim3{''}$. The flux densities determined for the compact source
at the center of NGC~6334A are given in Table 1.

\section{Results}

We found observations at 8.46 GHz for three epochs. Images from these observations
are shown in Figure 1. The compact source is clearly detected in the first two
epochs and an upper limit was set for the third epoch.
Clearly, the source is time variable, with a flux density in 1997 about twice as large as
in 2002. The upper limit for 2006 is not particularly stringent but it does indicate that
the source has not gone back to flux densities comparable to those of 1997.
For 1997 and 2006 the observations were centered on the nearby quasar
J17204-3554 (Rodr\'\i guez et al. 2012b)
and the flux densities have been corrected for the response of the primary beam.
The positions of the compact source determined for 1997 and 2002 differ by about
$0\rlap.{''}4$ in the north-south direction (see Figure 1). We have checked the accuracy
of the phase calibrators used and discarded the
possibility of attributing this effect to a poorly
determined position of these calibrators.
There is a possibility that we
are observing large proper motions but for the moment we will adopt the conservative
hypothesis that this difference is simply due to the low declination of the
source, that can introduce systematic shifts in position. We expect to test this
discrepancy in the future with new observations.

There is no counterpart to the compact radio source at other wavelengths, despite
sensitive searches in the optical (Russeil et al. 2012), infrared (Straw \& Hyland 1989)
and X-rays (Feigelson et al. 2009). Most probably, this is the result of the
large extinction expected toward this embedded source.

For 2002 we also have observations at 14.9 and 43.3 GHz. The source is marginally detected
at 14.9 GHz (see Table 1 and Figure 2) and an upper limit was set at 43.3 GHz.
By fitting a power-law to the 8.46 and 14.9 GHz detections, we obtain the
spectral index of the spectrum (see Figure 3), given by $S_\nu \propto \nu^{-1.2\pm0.5}$.
This spectral index is consistent with the values expected from optically-thin synchrotron
emission (i.e. Longair 2011). Our estimate of the spectral index was made assuming that
the source is unresolved at all frequencies observed.

\begin{figure}
\centering
\includegraphics[scale=0.4, angle=0]{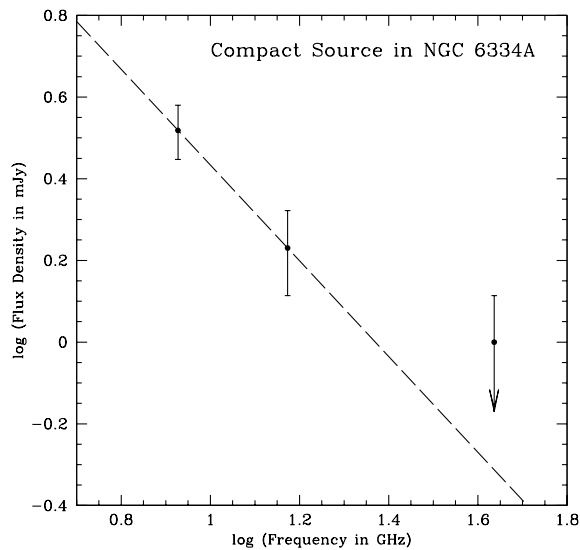}
 \caption{Spectrum of the continuum emission of the compact source in NGC~6334A.
 The dashed line is the $S_\nu \propto \nu^{-1.2}$ fit.
}
   \label{fig3}
\end{figure}

\section{Discussion and Conclusions}

Our results indicate that the compact radio continuum source at the center
of NGC~6334A is time variable on a scale of years and has a spectral index suggestive
of optically-thin synchrotron emission. The type of emission that best fits these characteristics
corresponds to that coming from
the wind interaction region of massive binary systems (see, for example, the
case of Cyg OB2 \#5 recently discussed by Ortiz-Le\'on et al. 2011 and Dzib et al. 2013b).
In these sources the winds of the components of the binary produce a shocked region between the
stars where electrons can reach relativistic speeds by Fermi acceleration,
producing synchrotron emission. As the binary system rotates, the wind interaction region
can move behind the optically thick free-free envelopes produced by the stellar winds and
become undetectable. This interpretation can be tested with future observations, made over
a long duration of time. In this type of sources the variability of the radio emission is
clearly periodic (i.e. Dougherty \& Williams 2000; 
Blomme et al. 2013). Furthermore, the synchrotron emitting region is very compact and
can be detected and studied at the milli-arcsecond scale, with Very Long Baseline Interferometry
observations (i. e. Ortiz-Le\'on et al. 2011; Dzib et al. 2013b). 

We are also undertaking
a reanalysis of good quality VLA observations of compact, ultracompact, and hypercompact
HII regions to search for additional examples. With only two cases
reported to date it is very difficult to reach
general conclusions. For example, the compact source associated with W3(OH) appears to be thermal,
while that associated with NGC~63334A and discussed here appears to be non thermal,
suggesting more than one origin for this type of source.

\adjustfinalcols

\acknowledgments
LFR, LL, and SEK are thankful for the support
of DGAPA, UNAM, and of CONACyT (M\'exico).
This research has made use of the SIMBAD database, 
operated at CDS, Strasbourg, France.

\vskip0.5cm


\end{document}